\documentclass[prd,aps,tightenlines,longbibliography,nofootinbib%
,12pt%
]{revtex4-1}
\usepackage[utf8]{inputenc} 
\usepackage{amssymb}
\usepackage{amsbsy}
\usepackage{chngcntr}
\usepackage{amsmath}
\usepackage{amsfonts}
\usepackage{verbatim}
\usepackage{graphicx}

\def\roughly#1{\mathrel{\raise.3ex
\hbox{$#1$\kern-.75em\lower1ex\hbox{$\sim$}}}}

\newcommand{\be}{\begin{equation}}

\newcommand{\ee}{\end{equation}}
\newcommand{\bqa}{\begin{eqnarray}}
\newcommand{\eqa}{\end{eqnarray}}

\newcommand{\hc}{\hat{c}}
\newcommand{\bea}{\begin{eqnarray}}
\newcommand{\eea}{\end{eqnarray}}

\newcommand{\ct}{\cos\theta}

\newcommand{\VB}{\mathbf{B}}
\newcommand{\VE}{\mathbf{E}}
\newcommand{\vf}{\mathbf{f}}
\newcommand{\vF}{\mathbf{F}}

\newcommand\Bxyz[3]{\left\{ \begin{array}{r}
                             #1 \\ #2 \\ #3
                            \end{array} \right\}}

\newcommand{\Lc}{\mathcal{L}^{(1)}}
\newcommand{\x}{{\cal F}}
\newcommand{\y}{{\cal G}}

\begin{document}

\rightline{BI-TP 2018/02}

\rightline{TUW-18-04}

\vspace{1.0cm}

\title{Light-by-Light Scattering in the Presence of Magnetic Fields}

\author{ R.~Baier} 
\email {baier@physik.uni-bielefeld.de}

\affiliation{Faculty of Physics, University of 
Bielefeld, D-33501 Bielefeld, Germany}

\author{A.~Rebhan} 
\email {anton.rebhan@tuwien.ac.at}

\author{M.~W\"odlinger}
\email{matthias.woedlinger@gmail.com}
\affiliation{Institute for Theoretical Physics, Technische Universität Wien, Wiedner Hauptstr.\ 8-10, A-1040 Vienna, Austria}


\begin{abstract}
\vspace{2cm}
The low-energy light-by-light cross section as determined by the nonlinear Euler-Heisenberg
QED Lagrangian is evaluated in the presence of constant magnetic fields {in the center-of-mass system
of the colliding photons}.
This cross section has a complicated dependence on directions and polarizations.
The overall magnitude decreases as the magnetic field is increased from zero,
but this trend is reversed for ultrastrong magnetic fields $B\gtrsim B_c$, where the cross section
eventually grows quadratically with the magnetic field strength perpendicular to the collision axis. This effect is
due to interactions involving the lowest Landau level of virtual Dirac particles; it is
absent in scalar QED. 
An even more dramatic effect is found for virtual charged vector mesons where the one-loop cross section
diverges at the critical field strength due to an instability of the lowest Landau level and the possibility
of the formation of a superconducting vacuum state.
We also discuss (the absence of) implications for the recent observation of light-by-light
scattering in heavy-ion collisions.
\end{abstract}

\maketitle

\section{Introduction}

Scattering of light by light is a prediction of quantum electrodynamics (QED) that has been first
calculated in 1935, in fact prior to the full development of QED, in the low-energy limit by Euler and Kockel \cite{Euler:1935zz,Euler:1936},
and in the ultrarelativistic limit shortly thereafter by Akhiezer, Landau, and Pomeranchuk \cite{Akiezer1936,LL4}.
The former calculations were extended by Heisenberg and Euler \cite{Heisenberg:1935qt} who
obtained an effective low-energy Lagrangian which includes background electromagnetic fields to all orders
in field strength
(for historical reviews and references see
\cite{Dunne:2004nc,Dunne:2012vv,Dittrich:2014bxa,Scharnhorst:2017wzh};
a short list of further relevant references with regard to applications in light-by-light scattering is given by \cite{Itzykson:1980rh,Bern:2001dg,Liang:2011sj,dEnterria:2013zqi,Klusek-Gawenda:2016euz,
Rebhan:2017zdx,Ellis:2017edi,Gies:2017ezf}).

In high-energy ultraperipheral  collisions of heavy ions (HIC)
evidence of the quantum mechanical process of light-by-light scattering has been presented for the first time by the ATLAS collaboration at the LHC
\cite{Aaboud:2017bwk}, and more recently also by the CMS collaboration
\cite{dEnterria:2018uly}.
 Light-by-light scattering can be studied through the large (almost) real photon fluxes available in ultraperipheral hadron-hadron, best in lead-lead collisions at LHC. 
 
 In the noncentral HICs very strong magnetic fields are created perpendicular to the heavy ion reaction plane, which, however, decay rapidly, but are still strong at collision time $\tau \simeq 1 \,\mathrm{fm}$. The field strength has been estimated to reach \cite{Kharzeev:2007jp,Bzdak:2011yy,Deng:2012pc,Itakura:2013cia}
\be
B/B_c (\tau = 0\,\mathrm{fm}) \simeq O(10^5)~~ \text{and}~~ B/B_c (\tau = 0.6\,\mathrm{fm}) 
\simeq O(10^2 \text{--} 10^3)~, 
\ee
at RHIC for impact
 parameters $b \simeq 10 \,\mathrm{fm}$,
 with the critical magnetic field
 $B_c = \frac{m_e^2}{e}\approx 0.86 \,{\rm MeV}^2\approx 4.4\times 10^{13}\,\mathrm{G}$ 
in terms of the electron mass $m_e$.
At the LHC the estimated initial value is about a factor of 10 higher (but decays faster).

Motivated by this, the present paper considers $\gamma +\gamma \rightarrow \gamma+\gamma$
scattering in the presence of weak and strong (constant) magnetic fields {in the center-of-mass system
of the colliding photons,
from $B/B_c\ll 1$ to  $B/B_c \gg 1$ (but, parametrically, $B/B_c \ll \alpha^{-1/2}$ so that higher-loop corrections as well as 
the effects from dispersion and refraction of light in the
magnetic field \cite{Adler:1971wn} remain negligible).} In the following this process will be studied in detail
in the low-energy approximation provided by the Euler-Heisenberg Lagrangian.
In this regime, the cross section rises proportional to $\omega^6/m^8$ with increasing photon energy $\omega$.
At $\omega\sim m$ the cross section reaches its maximum value $\propto \alpha^4/m^2$ and afterwards decays rapidly like $1/\omega^2$
\cite{Akiezer1936,Karplus:1950zz,Bern:2001dg} until the next heavier charged particle starts to contribute
according to the Euler-Heisenberg Lagrangian 
but with a maximum value
that is suppressed by the corresponding lower inverse mass squared. After electrons and muons, also scalar
charged particles such as pions and kaons contribute, which are described by a variant of the Euler-Heisenberg
Lagrangian first obtained by Weisskopf \cite{Weisskopf:1936}. Also working out the effects
of magnetic background fields on virtual scalars, we find that magnetic fields lead to a monotonic decrease of
the light-by-light scattering cross section in scalar QED, whereas the lowest Landau level of the Dirac spinors contributes a counteracting 
effect that dominates at large magnetic fields where it leads to a growing cross section.
A theoretically particularly interesting case is given by the Euler-Heisenberg Lagrangian for charged vector bosons
\cite{Vanyashin:1965ple} for which we find a light-by-light scattering cross section growing with magnetic field strength
and diverging at the critical magnetic field 
where it has been conjectured that a charged vector boson condensate may form \cite{Ambjorn:1988fx,Chernodub:2010qx,Hidaka:2012mz,Chernodub:2013uja}.

As discussed further in the concluding section, relatively more significant effects from magnetic fields
are to be expected for lighter particles as they have smaller critical $B_c = m^2/e$. At least sufficiently below the mass
threshold, where the cross section steeply rises with energy, the Euler-Heisenberg Lagrangian permits reliable calculations
of the effects of magnetic fields on light-by-light scattering.

\section{Effective Lagrangian}

The one-loop effective QED Lagrangian for a Dirac particle with charge $e$ and mass $m$
in the presence of electromagnetic background fields with negligible gradients 
as obtained first by Heisenberg and Euler
reads \cite{Heisenberg:1935qt,Schwinger:1951nm,Dittrich:2000zu}
\begin{eqnarray}
\Lc_{\text{spinor}}&=&-\frac{1}{8\pi^2}\!\!
  \int\limits_0^{\infty}
  \!\!\frac{ds}{s^3}\mathrm{e}^{\!-m^2\!s}\!~\biggl[(es)^2|\y|
  \coth\Bigl(\!es\bigl(\!
  \sqrt{\x^2\!+\!\y^2}\!+\x\bigr)\!^{\frac{1}{2}}\!\Bigr) \nonumber\\
&&\qquad\qquad\times\cot\Bigl(\! es\bigl(\!\sqrt{\x^2\!+\!\y^2}
  \!-\x\bigr)\!^{\frac{1}{2}}\!\Bigr)\!-\frac{2}{3}(es)^2
  \x-1\biggr],
\label{1}
\end{eqnarray}
where $\x$ and $\y$ denote the Lorentz scalar and pseudoscalar
\begin{eqnarray}
\x&:=&\frac{1}{4}F_{\mu\nu}F^{\mu\nu} =
  \frac{1}{2}({\VB}^2-{\VE}^2) \,, \label{5a}\\
\y&:=&\frac{1}{4}F_{\mu\nu}\, ^\star\! F^{\mu\nu} = 
  {\VE}\cdot {\VB}\, , \label{5b}
\end{eqnarray}
that can be built from the field-strength tensor and its dual,
\begin{eqnarray}
F^{\mu\nu}&=&\partial^\mu A^\nu-\partial^\nu A^\mu\, \label{4a}\\
^\star\! F^{\mu\nu}&=& \frac{1}{2} \epsilon^{\mu\nu\alpha\beta}
  F_{\alpha\beta}\, .\label{4b}
\end{eqnarray}
The Maxwell Lagrangian
is given by  ${\cal L}^{(0)}=-\x$. 

An equivalent version of $(\ref{1})$ is
\begin{equation}
\Lc_{\text{spinor}}= -\frac{1}{8\pi^2}\!\! \int\limits_0^{\infty}
  \!\!\frac{ds}{s^3}\mathrm{e}^{\!-m^2\!s}~\!\biggl[(es)^2 a b~
  \coth\Bigl(\!es a\!\Bigr) 
~\cot\Bigl(\! es b \!\Bigr)\! 
-\frac{1}{3}(es)^2 (a^2 - b^2)
  -1\biggr]\! ,  
\label{2c}
\end{equation}
where  new variables are introduced\footnote{Here we follow the conventions used in Ref.~\cite{Dittrich:2000zu,Schubert:2001he} which differ from the original work
of Heisenberg and Euler \cite{Heisenberg:1935qt} as well as the review \cite{Dunne:2004nc} in the notational reversal $a\leftrightarrow b$.}
\begin{eqnarray}
a:=\bigl(\sqrt{ \x^2+\y^2} +\x\bigr)^{\frac{1}{2}}\, &,&\,
  b:=\bigl(\sqrt{   \x^2+\y^2} -\x\bigr)^{\frac{1}{2}}\, ,\label{38a} \\
\Longrightarrow \qquad |\y|=ab\quad &,&\, \x=\frac{1}{2}(a^2-b^2)\,
  .\label{2b}
\end{eqnarray}

In terms of the variables $a$ and $b$, the low-energy one-loop effective Lagrangian of QED with Dirac spinors replaced by charged scalars
reads \cite{Weisskopf:1936}
\begin{equation}
\Lc_{\text{scalar}} = \frac{1}{16\pi^2}\!\! \int\limits_0^{\infty}
  \!\!\frac{ds}{s^3}\mathrm{e}^{\!-m^2\!s}~\!\biggl[\frac{(es)^2 a b}{\sinh\left(es a\right) 
\,\sin\left(es b\right)} 
+\frac{1}{6}(es)^2 (a^2 - b^2)
  -1\biggr]\! ,
\label{Lscalar}
\end{equation}
{where $m$ is now the mass of the charged scalar particle}.
This is of potential interest for elastic light-by-light scattering when the photon energy approaches the mass scale of pions.

The 
Euler-Heisenberg Lagrangian for massive charged vector fields has been obtained in Ref.~\cite{Vanyashin:1965ple}
for the case of a gyromagnetic factor $g=2$, which is carried by the electroweak $W^\pm$ gauge bosons and
(approximately) also by the $\rho$ meson \cite{Samsonov:2003hs,Djukanovic:2005ag}.
It reads
\be
\Lc_{\text{vector}} = 3 \Lc_{\text{scalar}} 
   +\frac{e^2}{4\pi^2}\!\! \int\limits_0^{\infty}
  \!\!\frac{ds}{s} \left[
  \mathrm{e}^{-im^2 s} a\left(b \frac{\sin\left(es a\right)}{\sinh\left(es b\right)}-a\right)
  -\mathrm{e}^{-m^2 s} b\left(a \frac{\sin\left(es b\right)}{\sinh\left(es a\right)}-b\right)
  \right],\quad
\label{Lchargedvector}
\ee
{where $m$ on the right hand side, including the term $3 \Lc_{\text{scalar}}$, is the mass of the charged vector particle}.

For hadronic scalar and vector mesons, the effective Lagrangians (\ref{Lscalar}) and (\ref{Lchargedvector}) apply as long
as they can be treated as pointlike particles, which should be the case at sufficiently large photon wavelength
and sufficiently large Larmor radius $r_q\propto m_q/(eB)$ of the quark constituents, compared to the mesons' charge radii.

In the limit of weak fields, the various Euler-Heisenberg Lagrangians have the form
\be
\Lc=c_1 \x^2 + c_2\, \y^2+\ldots~~,\label{8a}
\ee
with $c_{1,2}$ given in Table \ref{tabc12}. These lowest-order terms are sufficient to
obtain the cross section for low-energy light-by-light scattering with zero background fields \cite{Euler:1935zz}
(see Ref.~\cite{Rebhan:2017zdx} for detailed results including polarization effects);
in the following the corresponding calculations will be generalized to a constant
magnetic background field of arbitrary strength.

\section{Geometry and kinematics}

\begin{figure}[b]
\includegraphics
{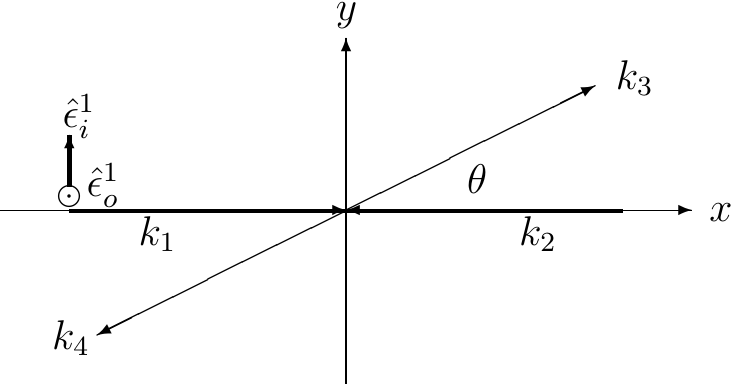}
\caption{Kinematics of photon-photon collisions in the center-of-mass system.\label{fig:kinematics}}
\end{figure}

The scattering amplitude $\cal M$ for $\gamma (k_1) + \gamma (k_2)
\rightarrow \gamma (k_3) + \gamma (k_4)$ is evaluated in the center-of-mass system,
\be
k_1 = (\omega, \omega {\hat k}) , ~ k_2 = (\omega, - \omega {\hat k}) ,
\nonumber 
\ee
\be
k_3 = (\omega, \omega {\hat k^\prime}) , ~ k_4 = (\omega, - \omega {\hat k^\prime}) .
\label{veck}
\ee
The scattering plane is defined by
\be
{\hat k} = (1, 0, 0) , ~ {\hat k^\prime} = (\cos\theta, \sin\theta, 0) .
\label{k2}
\ee

For linear polarizations the unit vectors ${\hat \epsilon}_i$ and
${\hat \epsilon}_o$ denote the directions in and out of the plane of scattering,
such that they form a right-handed orthogonal basis with the photon momenta
${\hat k}, {\hat k^\prime}$, respectively,
\bqa
{\hat \epsilon}_i^1 = (0, 1, 0)~, ~ {\hat \epsilon}_o^1 = ( 0, 0, 1) , 
\nonumber \\
{\hat \epsilon}_i^2 = (0, 1, 0)~, ~ {\hat \epsilon}_o^2 = ( 0, 0, -1) ,
\nonumber \\
{\hat \epsilon}_i^3 = (-\sin\theta, \cos\theta, 0)~,
 ~ {\hat \epsilon}_o^3 = ( 0, 0, 1) ,
\nonumber \\
{\hat \epsilon}_i^4 = (-\sin\theta, \cos\theta, 0)~,
 ~ {\hat \epsilon}_o^4 = ( 0, 0, -1) ~.
\label{pol}
\eqa
The radiation field strength vectors \cite{Adler:1971wn} are given by 
\bqa
\vf^{1 \pm}_{i,o} = \omega ({\hat k} \wedge {\hat \epsilon}_{i,o}^1 ~
\pm ~i~ {\hat \epsilon}_{i,o}^1)~,
\nonumber \\
\vf^{2 \pm}_{i,o} = \omega (- {\hat k} \wedge {\hat \epsilon}_{i,o}^2 ~
\pm ~i~ {\hat \epsilon}_{i,o}^2)~,
\nonumber \\
\vf^{3 \pm}_{i,o} = \omega ({{\hat k}^\prime} \wedge {\hat \epsilon}_{i,o}^3 ~
\pm ~i~ {\hat \epsilon}_{i,o}^3)~,
\nonumber \\
\vf^{4 \pm}_{i,o} = \omega (-{{\hat k}^\prime} \wedge {\hat \epsilon}_{i,o}^4 ~
\pm ~i~ {\hat \epsilon}_{i,o}^4)~.
\label{fs}
\eqa

The external fields are denoted by
\be
{\vF}^\pm = \VB \pm  ~i~ \VE
\label{ext}
\ee
with components ${F}^\pm_r,~  r=1,2,3$, as for the components $f^\pm_r$ of
$\vf^\pm$.

\section{Light-by-light scattering amplitudes and cross sections}

Following Adler's seminal work on photon splitting in a magnetic field \cite{Adler:1971wn} (as reviewed in Sect.~3.4 of Ref.~\cite{Dittrich:2000zu}), 
the matrix element for the scattering
$\gamma (k_1) + \gamma (k_2)
\rightarrow \gamma (k_3) + \gamma (k_4)$ 
in the presence of external electromagnetic fields is given by derivatives of the Euler-Heisenberg Lagrangian ({\ref{1}}) (or
its analogue (\ref{Lscalar}) in scalar QED and (\ref{Lchargedvector}) for charged vector mesons), which are finally evaluated for finite $\VB$ and vanishing $\VE = 0$.
The rather lengthy expression reads
\bqa
{\cal M} =
\Bigl(~f^{1 +}_r \cdot \frac{\partial}{\partial F^+_r} +
f^{1 -}_r \cdot \frac{\partial}{\partial F^-_r} \Bigr)
 \Bigl(~f^{2 +}_s \cdot \frac{\partial}{\partial F^+_s} +
f^{2 -}_ s\cdot \frac{\partial}{\partial F^-_s} \Bigr) 
\nonumber \\
\times ~\Bigl(~f^{3 +}_t \cdot \frac{\partial}{\partial F^+_t} +
f^{3 -}_t \cdot \frac{\partial}{\partial F^-_t} \Bigr)
 \Bigl(~f^{4 +}_u \cdot \frac{\partial}{\partial F^+_u} +
f^{4 -}_u \cdot \frac{\partial}{\partial F^-_u} \Bigr) \times \Lc~,
\label{mat1}
\eqa
and explicitly,
\be
{\cal M} =  f^{1 +}_r f^{2 +}_s f^{3 +}_t f^{4 +}_u 
~\frac{\partial^4 \Lc}{\partial{F^+_r}{\partial F^+_s}{\partial F^+_t}{\partial F^+_u}} 
\nonumber
\ee
\be
 ~+~\Bigl( f^{1 +}_r f^{2 +}_s f^{3 +}_t f^{4 -}_u
 +f^{1 +}_r f^{2 +}_s f^{4 +}_t f^{3 -}_u
+ f^{1 +}_r f^{3 +}_s f^{4 +}_t f^{2 -}_u 
 + f^{2 +}_r f^{3 +}_s f^{4 +}_t f^{1 -}_u \Bigr)
 ~\frac{\partial^4 \Lc}{\partial{F^+_r}{\partial F^+_s}{\partial F^+_t}{\partial F^-_u}}
\nonumber
\ee
\be
~+~\Bigl( f^{1 +}_r f^{2 +}_s f^{3 -}_t f^{4 -}_u
 +f^{1 +}_r f^{3 +}_s f^{2 -}_t f^{4-}_u
+ f^{1 +}_r f^{4 +}_s f^{2 -}_t f^{3 -}_u
 + f^{2 +}_r f^{3 +}_s f^{1 -}_t f^{4 -}_u 
\nonumber
\ee
\be
 ~+~ f^{3 +}_r f^{4 +}_s f^{1 -}_t f^{2 -}_u
 + f^{2 +}_r f^{4 +}_s f^{1 -}_t f^{3 -}_u \Bigr)
~ \frac{\partial^4 \Lc}{\partial{F^+_r}{\partial F^+_s}{\partial F^-_t}{\partial F^-_u}}
\nonumber
\ee
\be
~+~\Bigl(f^{1 -}_r f^{2 -}_s f^{3 -}_t f^{4 +}_u
 +f^{1 -}_r f^{2 -}_s f^{4 -}_t f^{3 +}_u
+ f^{1 -}_r f^{3 -}_s f^{4 -}_t f^{2 +}_u
 + f^{2 -}_r f^{3 -}_s f^{4 -}_t f^{1 +}_u \Bigr)
~ \frac{\partial^4 \Lc}{\partial{F^-_r}{\partial F^-_s}{\partial F^-_t}{\partial F^+_u}}
\nonumber
\ee
\be
+  f^{1 -}_r f^{2 -}_s f^{3 -}_t f^{4 -}_u
~\frac{\partial^4 \Lc}{\partial{F^-_r}{\partial  F^-_s}{\partial F^-_t}{\partial F^-_u}}~.
\label{matrix}
\ee

\noindent
Next the derivatives with respect to $F^\pm_r$ are expressed in terms of derivatives $\frac{\partial}{\partial \x}$ and $\frac{\partial}{\partial \y}$,
e.g.
\be
\frac{\partial}{\partial F^\pm_r} = \frac{1}{2} F^\pm_r
~(\frac{\partial}{\partial \x} \mp i~ \frac{\partial}{\partial \y})~,
\ee
using 
\be
\frac{\partial {\cal F}}{\partial F^\pm_r}=  \frac{1}{2} F^\pm_r~,~~
\frac{\partial {\cal G}}{\partial F^\pm_r}= \pm  \frac{1}{2 i} F^\pm_r~,
\ee
and
\be
\frac{\partial^2}{{\partial  F^+_r}{ \partial F^+_s}  } = 
\frac{1}{2} \delta_{rs} (\frac{\partial}{\partial \x} -~ 
i~ \frac{\partial}{\partial \y})~ +\frac{1}{4}
 F^+_r F^+_s~ (\frac{\partial^2}{\partial \x^2} -~ 
 2i~ \frac{\partial^2}{\partial \x \partial \y } -  \frac{\partial^2}{\partial \y^2})~, etc..
\ee
 
An important typical derivative is
\bqa
\frac{\partial ^4 \Lc}{\partial{F^+_r}{\partial F^+_s}{\partial F^+_t}{\partial F^+_u}} 
~=~\frac{1}{4} (\delta_{rs} \delta_{tu} + \delta_{rt} \delta_{su} + \delta_{st} \delta_{ru})
(\frac{\partial^2 \Lc}{\partial \x^2} - \frac{\partial^2 \Lc}{\partial \y^2})
\nonumber \\
~+~ \frac{1}{8} \Bigl(\delta_{rs} F^+_t F^+_s +\delta_{rt} F^+_s F^+_u +
\delta_{st} F^+_r F^+_u + \delta_{ru} F^+_s F^+_t + \delta_{su} F^+_r F^+_t
 +\delta_{tu} F^+_r F^+_s \Bigr) \nonumber \\
\times ~(\frac{\partial^3 \Lc}{\partial \x^3} - 
 3 \frac{\partial^3 \Lc}{ {\partial \x} {\partial \y^2}})
\nonumber \\
+ \frac{1}{16} F^+_r F^+_s F^+_t F^+_u ~
\Bigl(\frac{\partial^4 \Lc}{\partial \x^4} - 
 6 \frac{\partial^4 \Lc}{ {\partial \x^2} {\partial \y^2}}
+ \frac{\partial^4 \Lc}{  {\partial \y^4}} \Bigr)~, \nonumber \\
\label{examp}
\eqa
noting that odd derivatives with respect to $\y$ vanish for $\VE = 0$,
i.e. at $F^\pm_r = B_r$.


\subsection{Weak magnetic field}
\label{sec:weakfield}

In order to obtain the $O(\xi^2)$, $\xi = B/B_c$, correction to the leading-order matrix element ${\cal M}_{HE}$ of eq.(\ref{HE1}) the derivatives of Eq.~(\ref{der3}) enter, i.e.
\be
\delta {\cal M} =
 \frac{1}{8} ~{\cal M}_a ~ (\frac{\partial^3 \Lc}{\partial \x^3}
                 - 3 \frac{\partial^3 \Lc}{\partial \x \partial \y^2})
 ~+~  \frac{1}{8} ~{\cal M}_b ~ (\frac{\partial^3 \Lc}{\partial \x^3}
                 + \frac{\partial^3 \Lc}{\partial \x \partial \y^2})~,
\label{oxhi}
\ee
evaluated at $\x=\y=0$, where
\be
{\cal M}_a =
\nonumber
\ee
\be
(\vf^{1+} \cdot \vf^{4+}) (\vf^{2+} \cdot {\VB}) (\vf^{3+} \cdot {\VB})
~+~ (\vf^{2+} \cdot \vf^{4+}) (\vf^{1+} \cdot {\VB}) (\vf^{3+} \cdot {\VB})
~+~ (\vf^{3+} \cdot \vf^{4+}) (\vf^{1+} \cdot {\VB}) (\vf^{2+} \cdot {\VB})
\nonumber
\ee
\be
~+~ (\vf^{1+} \cdot \vf^{2+}) (\vf^{3+} \cdot {\VB}) (\vf^{4+} \cdot {\VB})
~+~ (\vf^{1+} \cdot \vf^{3+}) (\vf^{2+} \cdot {\VB}) (\vf^{4+} \cdot {\VB})
~+~ (\vf^{2+} \cdot \vf^{3+}) (\vf^{1+} \cdot {\VB}) (\vf^{4+} \cdot {\VB})
\nonumber
\ee
\be 
~+~ ( ~+~ \Longleftrightarrow ~-~ ) ~,
\label{MA1}
\ee
and
\be
{\cal M}_b =
\nonumber
\ee
\be
(\vf^{1+} \cdot \vf^{2+}) (\vf^{3+} \cdot {\VB}) (\vf^{4-} \cdot {\VB})
~+~ (\vf^{1+} \cdot \vf^{3+}) (\vf^{4+} \cdot {\VB}) (\vf^{2-} \cdot {\VB})
~+~ (\vf^{1+} \cdot \vf^{3+}) (\vf^{2+} \cdot {\VB}) (\vf^{4-} \cdot {\VB})
\nonumber
\ee
\be
~+~ (\vf^{1+} \cdot \vf^{2+}) (\vf^{4+} \cdot {\VB}) (\vf^{3-} \cdot {\VB})
~+~ (\vf^{2+} \cdot \vf^{3+}) (\vf^{4+} \cdot {\VB}) (\vf^{1-} \cdot {\VB})
~+~ (\vf^{1+} \cdot \vf^{4+}) (\vf^{2+} \cdot {\VB}) (\vf^{3-} \cdot {\VB})
\nonumber
\ee
\be
~+~ (\vf^{1+} \cdot \vf^{4+}) (\vf^{3+} \cdot {\VB}) (\vf^{2-} \cdot {\VB})
~+~ (\vf^{2+} \cdot \vf^{3+}) (\vf^{1+} \cdot {\VB}) (\vf^{4-} \cdot {\VB})
~+~ (\vf^{3+} \cdot \vf^{4+}) (\vf^{2+} \cdot {\VB}) (\vf^{1-} \cdot {\VB})
\nonumber
\ee
\be
~+~ (\vf^{2+} \cdot \vf^{4+}) (\vf^{3+} \cdot {\VB}) (\vf^{1-} \cdot {\VB})
~+~ (\vf^{2+} \cdot \vf^{4+}) (\vf^{1+} \cdot {\VB}) (\vf^{3-} \cdot {\VB})
~+~ (\vf^{3+} \cdot \vf^{4+}) (\vf^{1+} \cdot {\VB}) (\vf^{2-} \cdot {\VB})
\nonumber
\ee
\be
~+~ (\vf^{3-} \cdot \vf^{4-}) (\vf^{1+} \cdot {\VB}) (\vf^{2+} \cdot {\VB})
~+~ (\vf^{2-} \cdot \vf^{4-}) (\vf^{1+} \cdot {\VB}) (\vf^{3+} \cdot {\VB})
~+~ (\vf^{2-} \cdot \vf^{3-}) (\vf^{1+} \cdot {\VB}) (\vf^{4+} \cdot {\VB})
\nonumber
\ee
\be
~+~ (\vf^{1-} \cdot \vf^{4-}) (\vf^{2+} \cdot {\VB}) (\vf^{3+} \cdot {\VB})
~+~ (\vf^{1-} \cdot \vf^{2-}) (\vf^{3+} \cdot {\VB}) (\vf^{4+} \cdot {\VB})
~+~ (\vf^{1-} \cdot \vf^{3-}) (\vf^{2+} \cdot {\VB}) (\vf^{4+} \cdot {\VB})
\nonumber
\ee
\be 
~+~ ( ~+~ \Longleftrightarrow ~-~ ) ~.
\label{MB1}
\ee

With
\be
\Lc
=  c_1 \x^2 + c_2 \y^2 - {\hat c}_1 \frac{\x^3}{B_c^2} - {\hat c}_2
\frac{\x \y^2}{B_c^2} \pm ...,
\ee
and the explicit values derived in Appendix \ref{sec:weakLc} and tabulated in Table~\ref{tabc12},
the amplitudes for the linear polarizations in and out of the collision plane read\footnote{For vanishing
magnetic background fields, this agrees with the results given in Ref.~\cite{Rebhan:2017zdx} except that
a factor $-i$ has been absorbed in the definition of $\mathcal{M}$ as done also in Ref.~\cite{Adler:1971wn}.}

\bea 
\frac{\mathcal M_{oooo}}{\omega^4}&=&4 c_1(3+\cos^2\theta) + \Bxyz{-30\hc_1}{-30\hc_1}{-18\hc_1+16\hc_2}\xi^2
+\Bxyz{6\hc_1}{-42\hc_1}{-6\hc_1}\xi^2
\cos^2\theta,\\
\frac{\mathcal M_{iiii}}{\omega^4}&=&4 c_1(3+\cos^2\theta) + \Bxyz{-18\hc_1+4\hc_2}{-18\hc_1+4\hc_2}{-66\hc_1}\xi^2
+\Bxyz{-6\hc_1-4\hc_2}{-6\hc_1+12\hc_2}{-6\hc_1}\xi^2
\cos^2\theta,\\
\frac{\mathcal M_{ooii}}{\omega^4}&=& -8 c_1+4c_2 (1+  \cos^2 \theta) + \Bxyz{12\hc_1-6\hc_2}{24\hc_1-2\hc_2}{24\hc_1-14\hc_2}\xi^2
+\Bxyz{2\hc_2}{-14\hc_2}{-2\hc_2}\xi^2
\cos^2\theta,\\
\frac{\mathcal M_{iioo}}{\omega^4}&=& -8 c_1+4c_2 (1+  \cos^2 \theta) + \Bxyz{24\hc_1-2\hc_2}{12\hc_1-6\hc_2}{24\hc_1-14\hc_2}\xi^2
+\Bxyz{-12\hc_1-2\hc_2}{12\hc_1-10\hc_2}{-2\hc_2}\xi^2
\cos^2\theta,\qquad\\
\frac{\mathcal M_{oioi,ioio}}{\omega^4}&=&4~(c_1+c_2)(1+\ct)+2(c_2-c_1 )(3+\cos^2 \theta) + 
\Bxyz{3\hc_1-9\hc_2}{3\hc_1-9\hc_2}{9\hc_1-19\hc_2}\xi^2\nonumber\\
&&
+\Bxyz{-6\hc_1-2\hc_2}{-12\hc_1-4\hc_2}{-12\hc_1-4\hc_2}\xi^2\cos \theta
+\Bxyz{3(\hc_1+\hc_2)}{9\hc_1-11\hc_2}{3\hc_1-\hc_2}\xi^2
\cos^2\theta,\\
{\mathcal M_{oiio,iooi}}&=&{\mathcal M_{oioi,ioio}}\Big|_{\cos \theta \to - \cos \theta},
\eea
where the three entries within the curly brackets refer to $\VB$ pointing in $x$, $y$, and $z$ direction, respectively.
For such $\VB$, the remaining amplitudes with an odd number of $i$ or $o$ polarizations vanish identically.

\begin{table}[t]
\caption{\label{tabc12}Coefficients $c_{1,2}/C$ and $\hc_{1,2}/C$ with $C=\alpha^2/m^4$.}
\begin{ruledtabular}
\begin{tabular}{@{}ccccc@{}} 
 & $c_1/C$ & $c_2/C$ & $\hc_1/C$ & $\hc_2/C$ \\ 
 \colrule
spinor QED& 8/45 & 14/45 & 64/315 & 104/315\\
scalar QED& 7/90 & 1/90 & 31/315 & 11/315 \\
supersymmetric QED & 1/3 & 1/3 & 2/5 & 2/5\\
charged massive vector& 29/10 & 27/10 & $-137/105$ & $-157/105$\\ 
\end{tabular}
\end{ruledtabular}
\end{table}

While we refrain from listing the unwieldy general case of oblique orientations of the magnetic field
for all amplitudes,
Appendix \ref{sec:sigmaunpol} gives the general weak-field result for the resulting unpolarized cross section.
The resulting total unpolarized cross section reads
\bea
\sigma(\gamma\gamma\to\gamma\gamma)^{\rm unpol}&=&\frac12 \int d\Omega \frac{d\sigma^{\rm unpol}}{d\Omega}\nonumber\\&=&\frac{7(3c_1^2-2c_1c_2+3c_2^2)\omega^6}{20\pi}\nonumber\\
&&+\frac{\omega^6}{15\pi} \frac{B_\parallel^2}{B_c^2}(-57 c_1 \hc_1 + 18 \hc_1 c_2 + 10 c_1 \hc_2 - 23 c_2 \hc_2)\nonumber\\
&&+\frac{\omega^6}{120\pi} \frac{B_\perp^2}{B_c^2}(-717 c_1 \hc_1 + 243 \hc_1 c_2 + 233 c_1 \hc_2 - 391 c_2 \hc_2),
\eea
where $B_\parallel$ is the magnetic field component parallel to the collision axis of the photons and $B_\perp$ the part orthogonal to it.
For spinor QED this yields
\be
\sigma(\gamma\gamma\to\gamma\gamma)^{\rm unpol}_{\rm spinor}=\frac{973 \,\alpha^4 \omega^6}{10125\, \pi m^8}
\left[
1-\frac{38224 \,B_\parallel^2 + 65602 \,B_\perp^2}{20433~B_c^2}+O(\xi^4)
\right],
\ee
and for QED with a charged scalar field instead of a Dirac spinor one has
\be
\sigma(\gamma\gamma\to\gamma\gamma)^{\rm unpol}_{\rm scalar}=\frac{119\,\alpha^4 \omega^6}{20250\, \pi m^8}
\left[
1-\frac{11294 \,B_\parallel^2 +  16802\,B_\perp^2}{2499~B_c^2}+O(\xi^4)
\right].
\ee

Scalar QED is relevant for light-by-light scattering at energies below the peak in the cross section
produced by muons, since there charged pions also start to contribute. It is moreover particularly
interesting in that it highlights the effects of the magnetic moments in spinor QED:
In scalar QED, the total cross
section is only about 6\% of the result in spinor QED. (Even with two charged scalars so that
scalar QED has the same number of degrees of freedom, the cross section is less than a quarter of
that of spinor QED.)
This is reflected by the relatively small 
coefficients $c_2$ and $\hc_2$ associated with the terms involving the square of the pseudoscalar 
$\y=\frac{1}{4}F_{\mu\nu}\, ^\star\! F^{\mu\nu}$ (see Table~\ref{tabc12}).
Moreover, 
turning on a (subcritical) magnetic field decreases the total cross
section more than twice as strongly as is the case in spinor QED. In fact, as will be shown below,
the limit of strong magnetic fields is dominated by the lowest Landau level of Dirac spinors
which eventually
leads to an increase of the cross section.

As an aside we note that supersymmetric QED, which in Ref.~\cite{Rebhan:2017zdx} has been shown to have particularly
simple polarization patterns, gives the slightly simpler result 
\be
\sigma(\gamma\gamma\to\gamma\gamma)^{\rm unpol}_{\rm sQED}=\frac{7\,\alpha^4 \omega^6}{45\, \pi m^8}
\left[
1-\frac{104 \,B_\parallel^2 +  158 \,B_\perp^2}{35~B_c^2}+O(\xi^4)
\right].
\ee

Of potential interest to light-by-light scattering are also charged vector bosons, in particular at photon
energies between the pion and the $\rho$ meson mass scales. In hadronic contributions to light-by-light scattering, which is
a critical ingredient in calculations of the anomalous magnetic moment of muons \cite{Jegerlehner:2009ry}, 
it is usually assumed that at
the scale of the $\rho$ meson one can switch to quark degrees of freedom \cite{Bern:2001dg}. However, light-by-light scattering
through virtual quarks differs quite strongly from the one through virtual vector bosons. In Table \ref{tabc12} we have
also given the coefficients in the expansion of the Euler-Heisenberg Lagrangian resulting from vector mesons
with gyromagnetic factor $g=2$ \cite{Vanyashin:1965ple,Skalozub:1975ab,Preucil:2017wen} 
corresponding to nonabelian vector bosons as well as to vector mesons \cite{Djukanovic:2005ag}
(see also \cite{Samsonov:2003hs}). The interactions due to the magnetic moment of the vector
mesons turn out to have the effect of enhancing the light-by-light cross section already in the weak-field limit:
\be
\sigma(\gamma\gamma\to\gamma\gamma)^{\rm unpol}_{\rm vector}=\frac{2751\,\alpha^4 \omega^6}{250 \, \pi m^8}
\left[1 + \frac{211846 \,B_\parallel^2 + 318298 \,B_\perp^2}{173313~B_c^2}
+O(\xi^4)\right],
\ee
which is a stark difference to both scalar and spinor QED.
As we shall discuss presently, this difference becomes even more pronounced as $\xi$ approaches unity, where one
enters a regime with possible vector boson condensation \cite{Chernodub:2010qx,Hidaka:2012mz,Chernodub:2013uja}.
Furthermore, already at vanishing magnetic field, the total cross section for a charged vector boson is very much
larger than that produced by three scalar degrees of freedom of the same mass, to wit, by a factor of $3537/17\approx 208.06$,
underlining the importance of the magnetic moment of the virtual particles in light-by-light scattering.

\begin{figure}[b]
\includegraphics[width=0.6\textwidth]{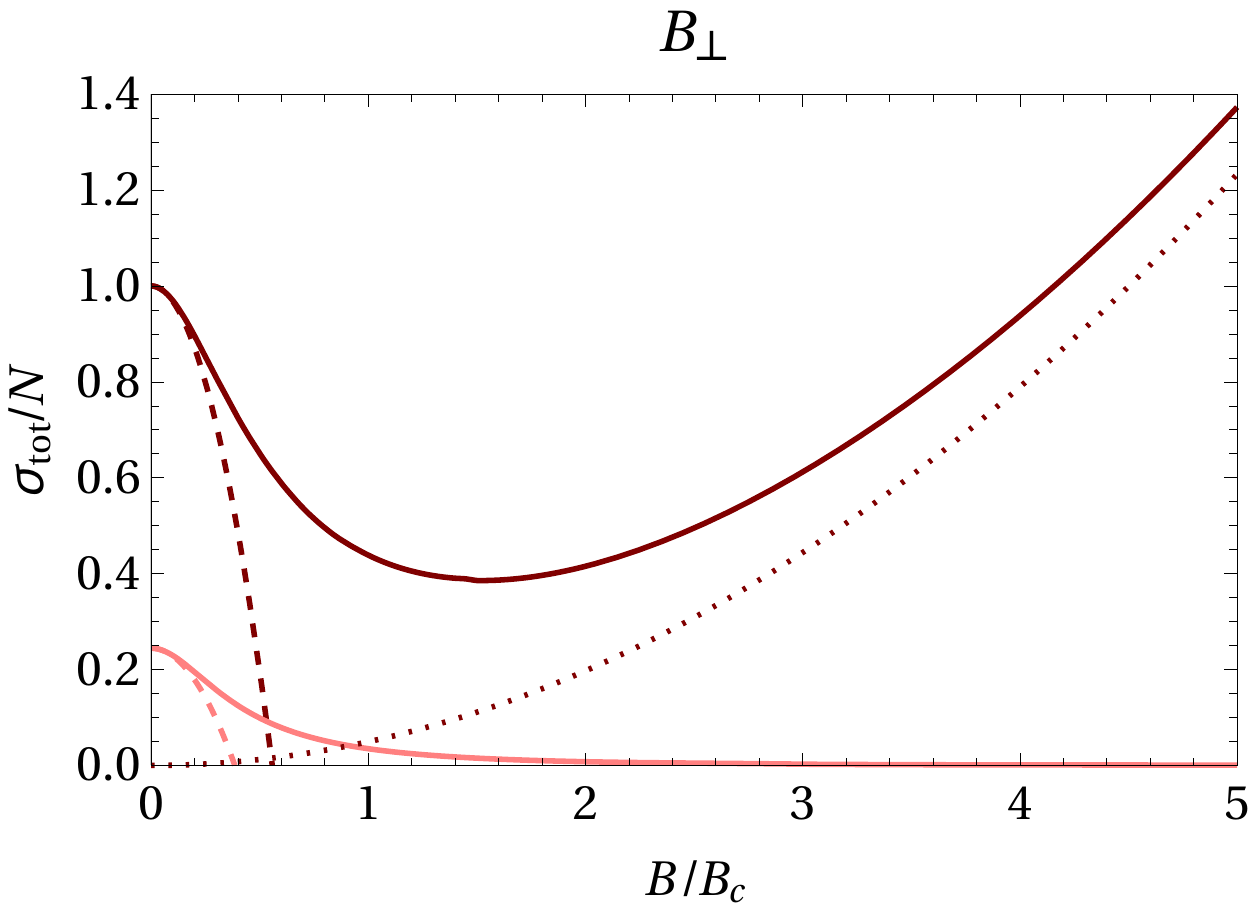}
\caption{\label{fig:stotBperp}Total cross section for unpolarized photons
as a function of $\xi=B/B_c$ with magnetic field perpendicular to the collision axis
for spinor QED (dark-red line) and for QED with two charged scalars (light-red line), both normalized to the total
cross section of spinor QED at zero magnetic field.
The weak-field result to order $\xi^2$ is given by the corresponding dashed lines. 
The strong-field result (\ref{stotlarge})
for spinor QED is given by the dotted black line.}
\end{figure}
\begin{figure}[t]
\includegraphics[width=0.6\textwidth]{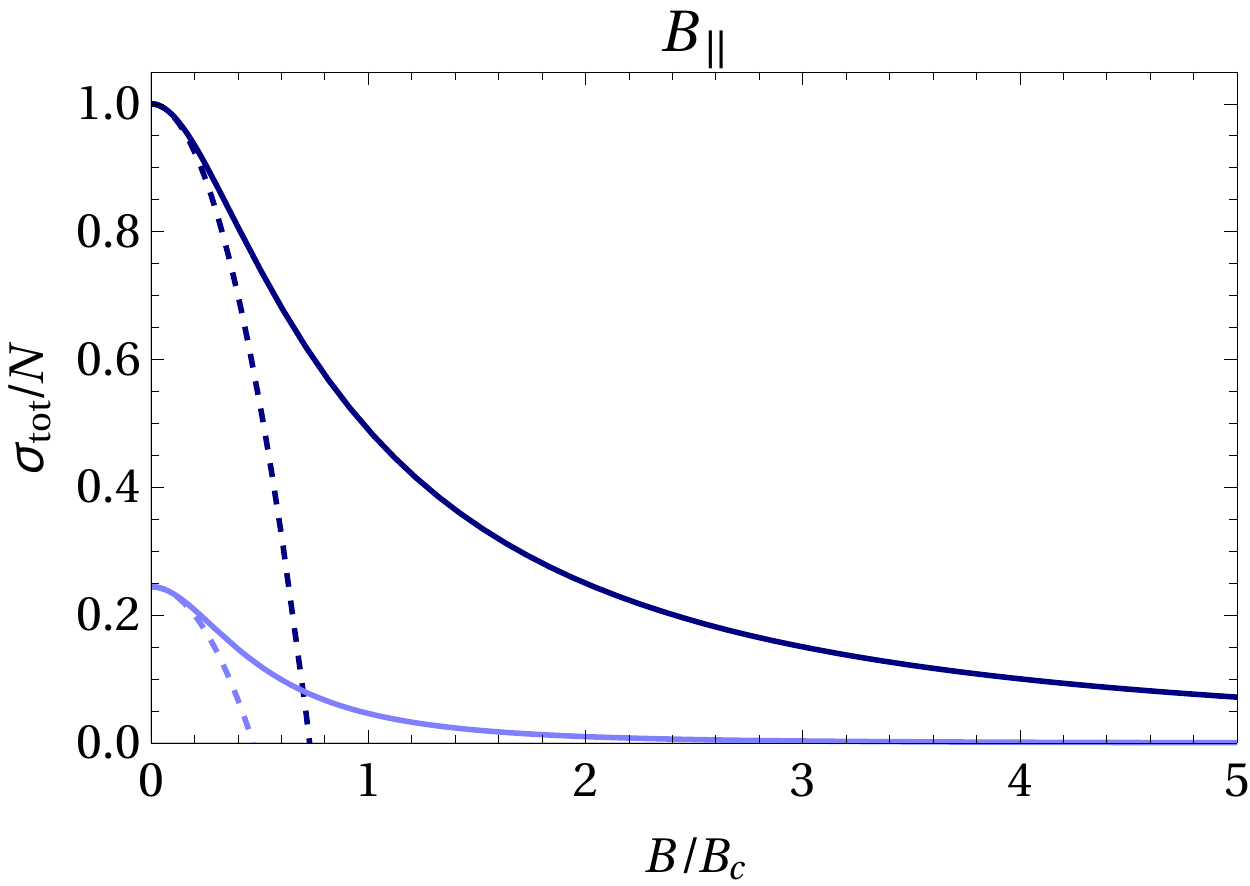}
\caption{\label{fig:stotBlong}Same as Fig.~\ref{fig:stotBperp} but with magnetic field parallel to the collision axis
(now dark-blue and light-blue coloring for spinor and scalar QED, respectively).}
\end{figure}

\subsection{Intermediate field strength}

For $\xi= {B}/{B_c} \gtrsim 0.5$, the weak-field expansion breaks down and one has to resort to numerical evaluations
of the integral representations of the various derivatives of $\mathcal{L}_c$ appearing in
(\ref{mat1}).

Our numerical results are shown in Fig.~\ref{fig:stotBperp} and \ref{fig:stotBlong} for 
magnetic fields perpendicular and parallel to the collision axis, respectively, where the former case
is the one of potential relevance to HIC.
In these plots we compare the result for spinor QED and scalar QED, where in the latter case two charged scalar
particles are assumed so that the difference between the two results is entirely due to the additional
interactions of the magnetic moment carried by Dirac spinors. Also given are the weak-field limits up to
order $\xi^2$ derived above, which are seen to become inaccurate around $\xi\simeq 0.5$.

For larger $\xi$, the results for scalar QED are seen to tend to zero rapidly ($\sim \xi^{-4}$ for $\xi\gg 1$),
whereas the spinor QED result for the case of perpendicular magnetic field has a minimum at $\xi\simeq 1.5$ after
which it grows quadratically with $\xi$. 

Further details that show up in differential cross sections are displayed in
Appendix \ref{sec:polardiagrams}.

\begin{figure}[h]
\includegraphics[width=0.6\textwidth]{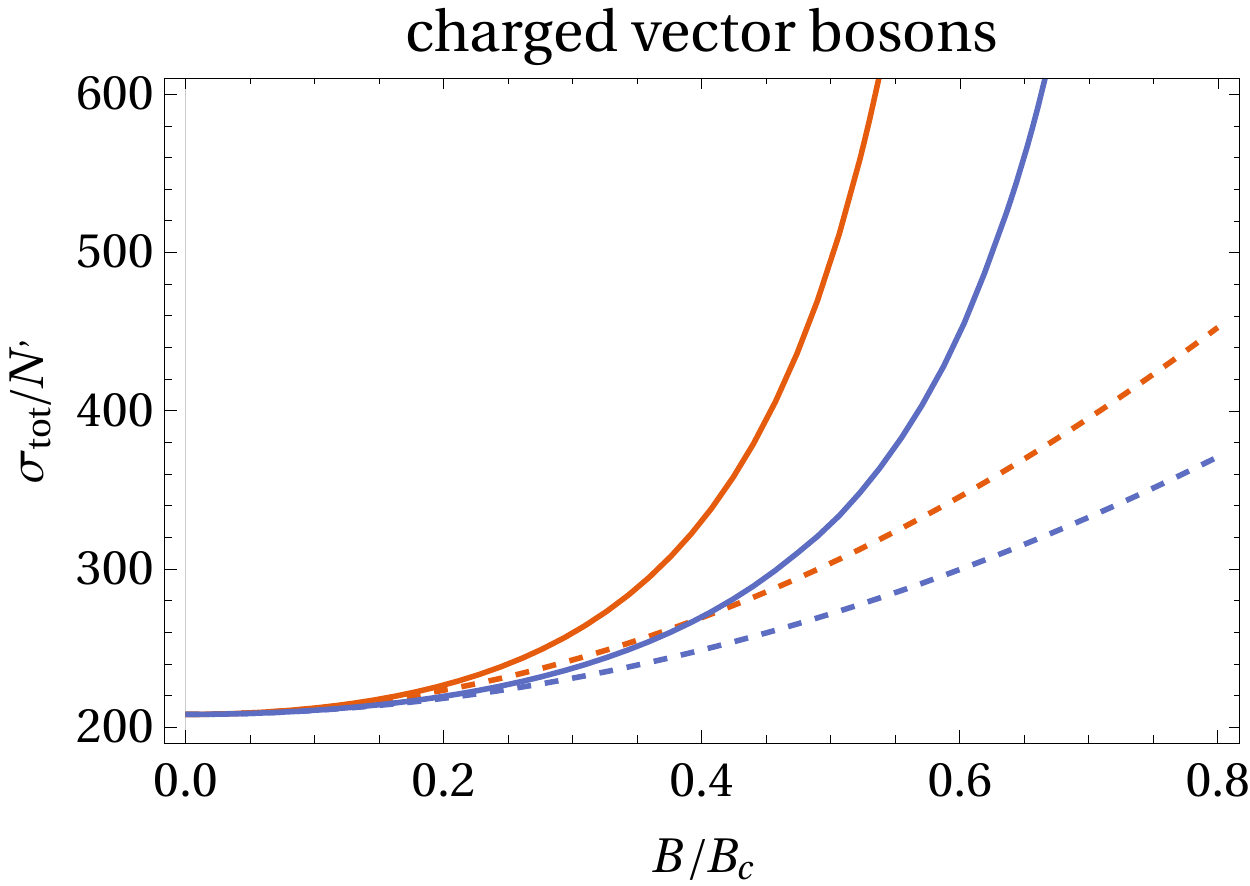}%
\caption{\label{fig:stotvector}
Total unpolarized light-by-light scattering cross section for virtual charged vector bosons with $g=2$
as a function of $\xi=B/B_c$
with the magnetic field perpendicular (red lines) and longitudinal (blue lines) to the collision axis (dashed lines
give the corresponding weak-field results). 
In order to highlight the effects of the magnetic moment of the charged vector bosons,
the normalization constant $N'$ is chosen as the $B=0$ result for three charged scalars of the same charge and mass,
which is a factor 
$3537/17\approx 208.06$ smaller than for one massive charged vector boson.}
\end{figure}

In the case of QED with charged vector bosons, for which the total cross section with magnetic field perpendicular or longitudinal
to the collision axis is evaluated in Fig.~\ref{fig:stotvector}, we find an increase which is quadratic in $\xi$ for small $\xi$
and which dramatically accelerates for larger $\xi$ with a divergence at $\xi=1$. 
In fact, at $\xi>1$ the lowest Landau level of a charged vector with $g=2$
becomes tachyonic, corresponding to the conjectured 
condensation of the charged vector bosons to form a superconducting vacuum \cite{Chernodub:2010qx,Hidaka:2012mz,Chernodub:2013uja}.
As explained in Appendix \ref{sec:strongfieldlimitd4Ldy4}, the calculation of the light-by-light scattering
cross section through the Euler-Heisenberg Lagrangian is
valid only for $\omega^2/m^2\ll 1-\xi$ so that the singularity is never reached.

\subsection{Strong magnetic field}

In the limit $\xi = {B}/{B_c} \gg1$ (but parametrically $\xi^2 \ll 1/\alpha$) the dominant contribution in spinor QED
comes from the derivative
${\partial^4 \Lc}/{\partial \y^4}$ at $\y=0$, so that e.g.
\be
\frac{\partial ^4 \Lc}{\partial{F^+_r}{\partial F^+_s}{\partial F^+_t}{\partial F^+_u}} 
\rightarrow 
\frac{1}{16} B_r B_s B_t B_u ~\frac{\partial^4\Lc}{\partial \y^4}\Big|_{\y=0}~.
\ee
Thus the matrix element in leading order of a strong magnetic field becomes
\bea
&&{{\cal M}}\big/\bigl(\frac{1}{16} \frac{\partial^4\Lc}{\partial \y^4}\bigr) =~
\nonumber
\\&&=(\vf^{1+} \cdot \VB)(\vf^{2+} \cdot \VB) (\vf^{3+} \cdot \VB)(\vf^{4+}
 \cdot \VB)
~+~(\vf^{1-} \cdot \VB)(\vf^{2-} \cdot \VB) (\vf^{3-} \cdot \VB)(\vf^{4-}
 \cdot \VB)
\nonumber
\\&&-~(\vf^{1+} \cdot \VB)(\vf^{2+} \cdot \VB) (\vf^{3+} \cdot \VB)(\vf^{4-}
 \cdot \VB)
~-~(\vf^{1+} \cdot \VB)(\vf^{2+} \cdot \VB) (\vf^{3-} \cdot \VB)(\vf^{4+}
 \cdot \VB)
\nonumber
\\&&-~(\vf^{1+} \cdot \VB)(\vf^{2-} \cdot \VB) (\vf^{3+} \cdot \VB)(\vf^{4+}
 \cdot \VB)
~-~(\vf^{1-} \cdot \VB)(\vf^{2+} \cdot \VB) (\vf^{3+} \cdot \VB)(\vf^{4+}
 \cdot \VB)
\nonumber
\\&&+~(\vf^{1+} \cdot \VB)(\vf^{2+} \cdot \VB) (\vf^{3-} \cdot \VB)(\vf^{4-}
 \cdot \VB)
~+~(\vf^{1+} \cdot \VB)(\vf^{2-} \cdot \VB) (\vf^{3+} \cdot \VB)(\vf^{4-}
 \cdot \VB)
\nonumber
\\&&+~(\vf^{1+} \cdot \VB)(\vf^{2-} \cdot \VB) (\vf^{3-} \cdot \VB)(\vf^{4+}
 \cdot \VB)
~+~(\vf^{1-} \cdot \VB)(\vf^{2+} \cdot \VB) (\vf^{3+} \cdot \VB)(\vf^{4-}
 \cdot \VB)
\nonumber
\\&&+~(\vf^{1-} \cdot \VB)(\vf^{2-} \cdot \VB) (\vf^{3+} \cdot \VB)(\vf^{4+}
 \cdot \VB)
~+~(\vf^{1-} \cdot \VB)(\vf^{2+} \cdot \VB) (\vf^{3-} \cdot \VB)(\vf^{4+}
 \cdot \VB)
 \nonumber
\\&&-~(\vf^{1-} \cdot \VB)(\vf^{2-} \cdot \VB) (\vf^{3-} \cdot \VB)(\vf^{4+}
 \cdot \VB)
~-~(\vf^{1-} \cdot \VB)(\vf^{2-} \cdot \VB) (\vf^{3+} \cdot \VB)(\vf^{4-}
 \cdot \VB)
\nonumber
\\&&-~(\vf^{1-} \cdot \VB)(\vf^{2+} \cdot \VB) (\vf^{3-} \cdot \VB)(\vf^{4-}
 \cdot \VB)
~-~(\vf^{1+} \cdot \VB)(\vf^{2-} \cdot \VB) (\vf^{3-} \cdot \VB)(\vf^{4-}
 \cdot \VB)~. 
\label{larg}
\eea

An amplitude with polarization vectors $\hat\epsilon^{1,2,3,4}$ (cf.\ Eq.~(\ref{pol})) is given
by
\be
{\cal M}=\omega^4 \frac{\partial^4 \Lc}{\partial^4 \y}\prod_{I=1}^4 \hat\epsilon^I\cdot \VB
= \frac{32 \alpha^2}{15} (\frac{\omega}{m})^4 ~\xi~ \prod_{I=1}^4 \hat\epsilon^I\cdot \hat \VB + O(\xi^0),
\label{MstrongB}
\ee
where $\hat \VB$ is the unit vector in the direction of $\VB$.
For example, when $\VB$ points in the $z$-direction, i.e., orthogonal to the scattering plane,
the only nonvanishing amplitude for linear polarizations is
\be
{\cal M}_{oooo}|_{B_x=B_y=0} = \frac{32 \alpha^2}{15} (\frac{\omega}{m})^4 ~\xi + O(\xi^0),
\ee
which is $\theta$-independent;
when $\VB$ points in the $y$-direction, i.e., in the scattering plane and orthogonal to the incoming photons,
the only nonvanishing amplitude is
\be
{\cal M}_{iiii}|_{B_x=B_z=0} = \frac{32 \alpha^2}{15} (\frac{\omega}{m})^4 ~\xi~\cos^2\theta + O(\xi^0),
\ee
which vanishes for outgoing photon momenta in the direction of $\VB$. 

The low-energy unpolarized  cross section averaged over initial and summed over final polarisations for $\xi \gg 1$
and arbitrary orientation of $\VB$ reads
\be
\frac{d \sigma^{\rm unpol}_{\rm spinor}}{d \Omega} = \frac{1}{(16 \pi)^2 \omega^2}
\frac{1}{4} ~\vert {\cal M} \vert^2
~=~ \frac{ \alpha^4 \omega^6}{225 \pi^2 m^8}~\xi^2~
\sin^4\beta ~\sin^4\beta',
\label{cross2}
\ee
where $\beta$ is the angle between $\VB$ and the direction of the incoming photon $\hat k$,
and $\beta'$ is the angle between $\VB$ and the outgoing direction $\hat k'$. Notice that
this differential cross section has the form of the square of a dipole radiation pattern,
with emission maximal in the plane orthogonal to the magnetic field.

The resulting unpolarized total cross section for $\xi \gg 1$ is
\be\label{stotlarge}
\sigma(\gamma\gamma\to\gamma\gamma)^{\rm unpol}_{\rm spinor}=\frac12 \int d\Omega \frac{d\sigma^{\rm unpol}}{d\Omega}
=\frac{ 16 \alpha^4 \omega^6}{3375 \pi m^8}~\xi^2~
\sin^4\beta .
\ee

As shown in Appendix \ref{sec:strongfieldlimitd4Ldy4}, the feature that for ultrastrong magnetic fields
the Euler-Heisenberg photon scattering cross section grows quadratically is absent in scalar QED.
It is entirely due to the magnetic moments of the virtual Dirac spinors which in the lowest Landau
level lead to a cancellation of magnetic interaction energy. 

\section{Discussion}

In this paper we have investigated the effect of sizable background magnetic fields on the light-by-light
scattering cross section in QED with charged scalar, spinor, or massive vector fields. 
We have found that the one-loop contribution of charged scalars
to the Euler-Heisenberg Lagrangian lead to a strong suppression of the light-by-light scattering cross section
for $B\gtrsim 0.5 B_c$. For spinor QED, the cross section initially also decreases with increasing magnetic field,
but this trend is reversed at $B\simeq 1.5 B_c$ after which the cross section grows quadratically with $B$.

Although at HIC the magnetic field reaches extremely large values with
respect to the critical one in terms of the electron mass $m_e$,
so that the light-by-light scattering
cross section would become correspondingly large, this applies
only at low photon energies
$\omega \lesssim m_e$. 

In the recent ATLAS measurement 
\cite{Aaboud:2017bwk} of light-by-light scattering the characteristic energy of the scattered photons
is in the range of several GeV, with peak values of the background magnetic field $B \sim 10^5\, \mathrm{MeV}^2$.
Because  the cross section decreases as $\alpha^4/\omega^2$ for $\omega\gg m$,
only massive loops can contribute effects due to external magnetic fields.
The critical magnetic field corresponding to the bottom and the charm quarks with mass $m_b\approx 4.2$~GeV and
$m_c\approx 1.25$~GeV
is $B_c(m_b)\sim 6\times 10^7\, \mathrm{MeV}^2$ and $B_c(m_c)\sim 5\times 10^6 \, \mathrm{MeV}^2$, respectively.
Effects from external magnetic fields at $\omega\lesssim m_b$ are therefore completely negligible.
For energies $\omega\lesssim m_c$, such effects would still be tiny;
noticeable effects on light-by-light scattering would seem to require
 photon energies $\omega\lesssim 0.1\,\mathrm{GeV}$, at or below the maximal contribution to the cross section from virtual muons
for which $B_c(m_\mu)\sim 4\times 10^4\, \mathrm{MeV}^2$.
However, with respect to the corresponding time scale $\omega^{-1}$, the magnetic field in HIC is then probably decaying too fast
to leave measurable effects.

A case of particular theoretical interest is that of charged $\rho$ mesons which have an unstable lowest Landau level
at $B\ge B_c(m_\rho)\sim 2\times 10^6\, \mathrm{MeV}^2$, where a superconducting vacuum state formed by
a condensate of $\rho^\pm$ mesons has been conjectured to arise \cite{Chernodub:2010qx}.\footnote{Evidence
in favor of this scenario from lattice gauge theory 
has been presented in \cite{Buividovich:2010tn,Braguta:2011hq}; see however \cite{Luschevskaya:2016epp,Bali:2017ian}.} 
In this paper we have also determined the contribution of charged vector mesons to light-by-light
scattering for photon energies $\omega\lesssim m_\rho$ as determined by the corresponding
Euler-Heisenberg Lagrangian derived in \cite{Vanyashin:1965ple}. 
This turns out to be enhanced by relatively large numerical prefactors
compared to scalar and spinor loops. Moreover, the cross section grows as the magnetic field strength is increased from zero.
Unfortunately, even the peak values of the magnetic field reached in HIC would give only effects below the percent level
to light-by-light scattering cross sections from virtual $\rho$ mesons (if the latter are included at all despite the
large width of the $\rho$ meson).

\begin{acknowledgments}
The authors would like to thank Maxim Chernodub, Dima Kharzeev, Massimiliano Procura, and Vladimir Skalozub for useful discussions of
the case of charged vector mesons.
\end{acknowledgments}

\appendix
\section{Matrix element for $\VB =0$}

For completeness the matrix element for $\gamma (k_1) + \gamma (k_2)
\rightarrow \gamma (k_3) + \gamma (k_4)$ for vanishing external fields
is given to fix the notation (see e.g. \cite{Itzykson:1980rh} and 
\cite{Bern:2001dg}),
\be
{\cal M} = \frac{c_1}{2} {\cal M}_1 - \frac{c_2}{2} {\cal M}_2 ,
\label{HE1}
\ee
\be
{\cal M}_1 = 
(\vf^{1+} \cdot \vf^{2+}) (\vf^{3+} \cdot \vf^{4+})  
~+~(\vf^{1+} \cdot \vf^{3+}) (\vf^{2+} \cdot \vf^{4+}) 
~+~(\vf^{2+} \cdot \vf^{3+}) (\vf^{1+} \cdot \vf^{4+}) 
\nonumber
\ee
\be
~+~(\vf^{1-} \cdot \vf^{2-}) (\vf^{3-} \cdot \vf^{4-})  
~+~(\vf^{1-} \cdot \vf^{3-}) (\vf^{2-} \cdot \vf^{4-}) 
~+~(\vf^{2-} \cdot \vf^{3-}) (\vf^{1-} \cdot \vf^{4-}) 
\nonumber
\ee
\be
~+~(\vf^{1+} \cdot \vf^{2+}) (\vf^{3-} \cdot \vf^{4-}) 
~+~(\vf^{1+} \cdot \vf^{3+}) (\vf^{2-} \cdot \vf^{4-}) 
~+~(\vf^{1+} \cdot \vf^{4+}) (\vf^{2-} \cdot \vf^{3-}) 
\nonumber
\ee
\be
~+~(\vf^{2+} \cdot \vf^{3+}) (\vf^{1-} \cdot \vf^{4-})  
~+~(\vf^{3+} \cdot \vf^{4+}) (\vf^{1-} \cdot \vf^{2-}) 
~+~(\vf^{2+} \cdot \vf^{4+}) (\vf^{1-} \cdot \vf^{3-}) ~~,
\nonumber
\label{M1}
\ee
and

\be
{\cal M}_2 = 
(\vf^{1+} \cdot \vf^{2+}) (\vf^{3+} \cdot \vf^{4+})  
~+~(\vf^{1+} \cdot \vf^{3+}) (\vf^{2+} \cdot \vf^{4+}) 
~+~(\vf^{2+} \cdot \vf^{3+}) (\vf^{1+} \cdot \vf^{4+}) 
\nonumber
\ee
\be
~+~(\vf^{1-} \cdot \vf^{2-}) (\vf^{3-} \cdot \vf^{4-})  
~+~(\vf^{1-} \cdot \vf^{3-}) (\vf^{2-} \cdot \vf^{4-}) 
~+~(\vf^{2-} \cdot \vf^{3-}) (\vf^{1-} \cdot \vf^{4-}) 
\nonumber
\ee
\be
~-~(\vf^{1+} \cdot \vf^{2+}) (\vf^{3-} \cdot \vf^{4-}) 
~-~(\vf^{1+} \cdot \vf^{3+}) (\vf^{2-} \cdot \vf^{4-}) 
~-~(\vf^{1+} \cdot \vf^{4+}) (\vf^{2-} \cdot \vf^{3-}) 
\nonumber
\ee
\be
~-~(\vf^{2+} \cdot \vf^{3+}) (\vf^{1-} \cdot \vf^{4-})  
~-~(\vf^{3+} \cdot \vf^{4+}) (\vf^{1-} \cdot \vf^{2-}) 
~-~(\vf^{2+} \cdot \vf^{4+}) (\vf^{1-} \cdot \vf^{3-}) ~~.
\nonumber
\label{M2}
\ee
For comparison the cross section is quoted (for references and a detailed evaluation see
 \cite{Rebhan:2017zdx}). For low energies $\omega \le m$ it is
\be
\frac{d \sigma^{\rm unpol}}{d \Omega} = \frac{\omega^6}{64 \pi^2}
(3 c_1^2 -  2 c_1 c_2 + 3 c_2^2) (3 + \cos^2\theta)^2~.
\label{crosQ}
\ee
In the high energy limit it decreases like
\be
\frac{d \sigma^{\rm unpol}}{d \Omega} \sim \frac{\alpha^4}{\omega^2}~,
\label{asymp}
\ee
beyond its maximum at $\omega \simeq 1.5 m$ \cite{LL4}. 

\section{Expansions for weak and strong background fields}
\subsection{Weak-field limit of $\Lc$}
\label{sec:weakLc}

The weak-field limit of the Euler-Heisenberg 
Lagrangian for spinors and scalars, Eqs.\ (\ref{2c}) and (\ref{Lscalar}), respectively, up to 
order $\xi^2 = (\frac{B}{B_c})^2$ is obtained by starting with the Taylor expansion for
\be
\coth z = \frac{1}{z} + \frac{z}{3} - \frac{z^3}{45} \pm \ldots,\quad 
1/\sinh z = \frac{1}{z} - \frac{z}{6} + \frac{7z^3}{360} \mp \ldots,
\ee
and
\be
\cot z = \frac{1}{z} - \frac{z}{3} - \frac{z^3}{45} - \ldots, \quad
1/\sin z = \frac{1}{z} + \frac{z}{6} + \frac{7z^3}{360} + \ldots,
\ee
leading in terms of the variables $a$ and $b$ to
\bqa
\Lc_{\text{spinor}}\!=\frac{e^4}{8\pi^2}\!\!
  \int\limits_0^{\infty}
  \!\!{ds}~{s}&&~\mathrm{e}^{\!-m^2\!s}\!~\biggl[
~\frac{a^4 + 5 a^2 b^2 + b^4}{45} \nonumber \\
 && - (es)^2 ~~\frac{2 a^6 + 7 a^4 b^2 - 7 a^2 b^4 - 2b^6}{945}
    ~~\biggr]\!  \pm \ldots,  \label{AA1}\\
\Lc_{\text{scalar}}\!=\frac{e^4}{16\pi^2}\!\!
  \int\limits_0^{\infty}
  \!\!{ds}~{s}&&~\mathrm{e}^{\!-m^2\!s}\!~\biggl[
~\frac{7a^4 -10 a^2 b^2 + 7b^4}{360} \nonumber \\
 && - (es)^2 ~~\frac{31 a^6 -49 a^4 b^2 +49 a^2 b^4 - 31 b^6}{15120}
    ~~\biggr]\!  \pm \ldots.  \label{AA1sc}
\eqa

For $a < m^2$ one can write the Euler-Heisenberg Lagrangian for charged vector bosons (\ref{Lchargedvector}) also as
\bqa
\Lc_{\text{vector}} &=& -\frac{1}{16\pi^2}\!\! \int\limits_0^{\infty}
  \!\!\frac{ds}{s^3}\mathrm{e}^{\!-m^2\!s}~\!\biggl[(es)^2 a b\frac{1-2\cosh(2esa)-2\cos(2esb)}{\sinh\left(es a\right) 
\,\sin\left(es b\right)} 
+\frac{7}{2}(es)^2 (a^2 - b^2)
  +3\biggr]\nonumber\\
  &=&\frac{e^4}{16\pi^2}\!\!
  \int\limits_0^{\infty}
  \!\!{ds}~{s}~\mathrm{e}^{\!-m^2\!s}\!~\biggl[
~\frac{29a^4 +50 a^2 b^2 + 29b^4}{40} \nonumber \\
 &&\qquad\qquad
 + (es)^2 ~~\frac{137 a^6 +217 a^4 b^2 -217 a^2 b^4 - 137 b^6}{15120}
    ~~\biggr]\!  \pm \ldots. 
\label{Lvectorsmall}
\eqa

After performing the $s$-integration one obtains
\be
\Lc =  c_1 \x^2 + c_2 \y^2 - {\hat c}_1 \frac{\x^3}{B_c^2} - {\hat c}_2
\frac{\x \y^2}{B_c^2} + \ldots,
\ee
in terms of the variables $\x$ and $\y$ (cf.~Eq.(\ref{2b})), with coefficients
as given in Table \ref{tabc12}.\footnote{The results obtained for charged vector mesons
are in agreement with those given in \cite{Liao:2012tj}.} 
(Supersymmetric QED has  $\Lc_{\mathrm{sQED}}=\Lc_{\mathrm{spinor}}+2\Lc_{\mathrm{scalar}}$.)

The contributions to the light-by-light scattering amplitudes
to order $\xi^2$ are obtained with $\x \rightarrow \frac{B^2}{2}$ and $\y \rightarrow 0$ from
\bea
\frac{\partial^2 \Lc}{\partial \x^2} = 2 c_1 - 3 {\hat{c}_1}\xi^2
+ \ldots,\qquad
\frac{\partial^2 \Lc}{\partial \y^2} = 2 c_2 - {\hat{c}_2}\xi^2
+ \ldots, \nonumber\\
\frac{\partial^3 \Lc}{\partial \x^3} =  
 -\frac{6 {\hat{c}_1}}{B_c^2} + \ldots,\qquad
~~ \frac{\partial^3 \Lc}{\partial \x \, \partial \y^2} =  
-\frac{2 {\hat{c}_2}}{B_c^2} + \ldots\,.
\label{der3}
\eea

\subsection{General expression for the unpolarized cross section to order $\xi^2$}
\label{sec:sigmaunpol}

In Sect.~\ref{sec:weakfield} the weak-field limit of the scattering amplitudes for linearly polarized photons has been
given for three cases of the orientation of the magnetic background field. The case of general orientation
is rather unwieldy for the various polarized cross sections, 
but a comparatively compact expression is obtained for the unpolarized cross section, which reads
\bea
\frac{d \sigma^{\rm unpol}}{d \Omega}&=&
\frac{\omega^6}{256\pi^2}(3 c_1^2 - 2 c_1 c_2 + 3 c_2^2) (7 + \cos 2 \theta)^2\nonumber\\
&&+\frac{\omega^6}{512\pi^2 B_c^2}\biggl\{
B_x^2 \Bigl[-1017 c_1 \hc_1 + 327 \hc_1 c_2 + 161 c_1 \hc_2 - 391 c_2 \hc_2 \nonumber\\
&&\qquad\qquad\qquad + 
   4 c_2 (15 \hc_1 + \hc_2) \cos2 \theta - 
   4 c_1 (33 \hc_1 + 7 \hc_2) \cos2 \theta \nonumber\\&&\qquad\qquad\qquad
   - 3 c_2 (\hc_1 - \hc_2) \cos4 \theta - 
   c_1 (3 \hc_1 + 5 \hc_2) \cos4 \theta\Bigr]\nonumber\\&&\qquad
+B_y^2 \Bigl[-1563 c_1 \hc_1 + 501 \hc_1 c_2 + 459 c_1 \hc_2 - 813 c_2 \hc_2 \nonumber\\&&\qquad\qquad- 
    4 c_1 (177 \hc_1 - 73 \hc_2) \cos2 \theta + 
    12 c_2 (21 \hc_1 - 37 \hc_2) \cos2 \theta \nonumber\\&&\qquad\qquad+ 
    c_2 (15 \hc_1 - 23 \hc_2) \cos4 \theta - 
    c_1 (33 \hc_1 - 17 \hc_2) \cos4 \theta\Bigr] \nonumber\\&&\qquad
+B_z^2 \Bigl[-1875 c_1 \hc_1 + 657 \hc_1 c_2 + 667 c_1 \hc_2 - 1073 c_2 \hc_2 \nonumber\\&&\qquad\qquad- 
    420 c_1 \hc_1 \cos2 \theta + 
    12 c_2 (9 \hc_1 - 17 \hc_2) \cos2 \theta + 
    100 c_1 \hc_2 \cos2 \theta \nonumber\\&&\qquad\qquad- 9 c_1 \hc_1 \cos4 \theta + 
    3 c_2 (\hc_1 - \hc_2) \cos4 \theta + c_1 \hc_2 \cos4 \theta\Bigr] \nonumber\\&&\qquad
+B_x B_y \Bigl[(516 c_1 \hc_1  - 156 \hc_1 c_2  - 
   276 c_1 \hc_2  + 396 c_2 \hc_2 ) \sin2 \theta \nonumber\\&&\qquad\qquad+ 
   (30 c_1 \hc_1  - 18 \hc_1 c_2  - 
   22 c_1 \hc_2  + 26 c_2 \hc_2 ) \sin4 \theta\Bigr]
\biggr\} + O(\xi^4).
   \eea

For the particularly important case of spinor QED this yields
\bea
\frac{d \sigma^{\rm unpol}_{\rm spinor}}{d \Omega}&=&\frac{\alpha^4 \omega^6}{64(45 \pi)^2 m^8}
\biggl\{ 139 (7 + \cos 2 \theta)^2\nonumber\\&&
-\frac{2}{7} \frac{B_x^2}{B_c^2} (41441 + 1956 \cos2 \theta + 251 \cos4 \theta)\nonumber\\&&
-\frac{2}{7} \frac{B_y^2}{B_c^2} (72075 + 33764 \cos2 \theta + 1425 \cos4 \theta)
\nonumber\\&&
-\frac{2}{7} \frac{B_z^2}{B_c^2} (86167 + 20756 \cos2 \theta + 341 \cos4 \theta)\nonumber\\&&
+\frac{4}{7} \frac{B_x B_y}{B_c^2} (14730 \sin2 \theta + 587 \sin4 \theta)\biggr\}+O(\xi^4).
\eea

\subsection{Strong-field limit of ${\partial^4 \Lc}/{\partial {\y^4}}$}
\label{sec:strongfieldlimitd4Ldy4}

In spinor QED,
the asymptotic behavior for $\xi = \frac{B}{B_c} \gg 1$ and $\y
\rightarrow 0$  is determined by the terms in the integrand of the Euler-Heisenberg 
Lagrangian (\ref{1})  proportional to

\begin{equation}
\label{A1}
\coth(es (\sqrt{\x^2 + \y^2} +\x)^{1/2}) = \coth(esa) \rightarrow \coth{t} ~,
\end{equation}
with $t = esa \rightarrow esB$.

Performing the Taylor expansion
\begin{eqnarray}
\y \cot(esb) = \frac{\y}{esb} ~ [ 1 - \frac{(esb)^2}{3} - \frac{(esb)^4}{45} + ..~ ] 
   \nonumber \\
\simeq \frac{B}{es} ~[ 1 - \frac{(es)^2}{3 B^2} \y^2 - \frac{(es)^4}{45 B^4} \y^4 + .. ~] ~,
\label{A2}
\end{eqnarray}
for $\y \rightarrow 0$ with $b \simeq \y/B , ~ esb \simeq es\y/B $,
one obtains
\begin{eqnarray}
 \frac{\partial^4 \Lc_{\text{spinor}}}{\partial {\cal{G}}^4}
\simeq \frac{1}{8 \pi^2} \frac{24}{45 B^3} \int_0^\infty \frac{ds}{s^3} (es)^5~
\mathrm{e}^{\!-m^2\!s}~ \coth(esB)  \nonumber \\
\simeq \frac{e^2}{8 \pi^2 B^6} \frac{8}{15}
 \int_0^\infty 
 ~dt~\mathrm{e}^{\!-t/\xi}~ t^2 \coth{t} ~,
\label{A4}
\end{eqnarray}
i.e. asymptotically for $\xi \gg 1$,
\begin{equation}
\frac{\partial^4 \Lc_{\text{spinor}}}{\partial {\y^4}}
\simeq \frac{e^2}{8 \pi^2 B^6} \frac{16}{15}~ \xi^3 ~~,
\label{A5}
\end{equation}
in agreement with the results derived in \cite{Heyl:1996dt}.
Since in the scattering amplitude (\ref{MstrongB}) this is combined with four powers of the magnetic field,
one has $\mathcal{M}\propto B$ in the limit of ultrastrong fields.
  
This result is, however, a special feature of spinor QED.
The Euler-Heisenberg Lagrangian for scalar QED (\ref{Lscalar}) as obtained originally by Weisskopf \cite{Weisskopf:1936} differs 
by the absence of the interaction term $\frac{e}{2} \sigma_{\mu\nu} F^{\mu\nu}$.
This has the effect that instead of
the functions $\coth(esa)$ and $\cot(esb)$ in (\ref{1}) one has $1/\sinh(esa)$ and $1/\sin(esb)$ \cite{Dunne:2004nc}.
In place of (\ref{A4}) one obtains
\be
\int_0^\infty 
 ~dt~\mathrm{e}^{\!-t/\xi}~ t^2 /\sinh{t}=\frac72\zeta(3)+O(\xi^{-1})\approx 4.207\ldots+O(\xi^{-1})
\ee
in the large-$\xi$ limit. This leads to contributions to $\cal M$ that are suppressed $\propto B^{-2}$ at large $B$.

As is particularly clear in the derivation of the Euler-Heisenberg Lagrangian due to Schwinger \cite{Schwinger:1951nm}, the
interaction with a spin magnetic moment $g\mu_B/2$ contributes the factor $\cosh(gesa/2)\cos(gesb/2)$, which for
$g=2$ compensates the exponential decay of $1/\sinh(esa)$, corresponding to the fact that then the magnetic interaction energy
of a Dirac spinor cancels in the lowest Landau level. This in fact suggests that also for Dirac spinors the rise of the photon-photon
scattering amplitude $\sim\xi$ 
will be modified eventually by higher-order effects at $\xi\gtrsim \alpha^{-1}$, when $(g-2)eB\gtrsim m^2$. 
However, already at the parametrically smaller order $\xi\gtrsim \alpha^{-1/2}$ our calculations would need to be
modified by including dispersion effects from nontrivial indices of refraction and birefringence \cite{Adler:1971wn}.

In the case of the Euler-Heisenberg Lagrangian for charged vector bosons with $g=2$ obtained in \cite{Vanyashin:1965ple}
the effects of the magnetic moment at high magnetic fields are even more dramatic.
The magnetic interaction energy, which leads to a modified mass
\be
m^2\to m^2_{\rm eff}=m^2+(2n-gs_z+1)eB,\quad n\ge0,
\ee
for spin projection $s_z$ along the magnetic field,
now reduces the effective mass of the lowest Landau level,
such that it becomes imaginary for $eB>m^2$,
corresponding to the potential instability of the vacuum against formation of a superconducting condensate of
charged vector bosons \cite{Chernodub:2010qx}.

\begin{figure}
\includegraphics[width=0.6\textwidth]{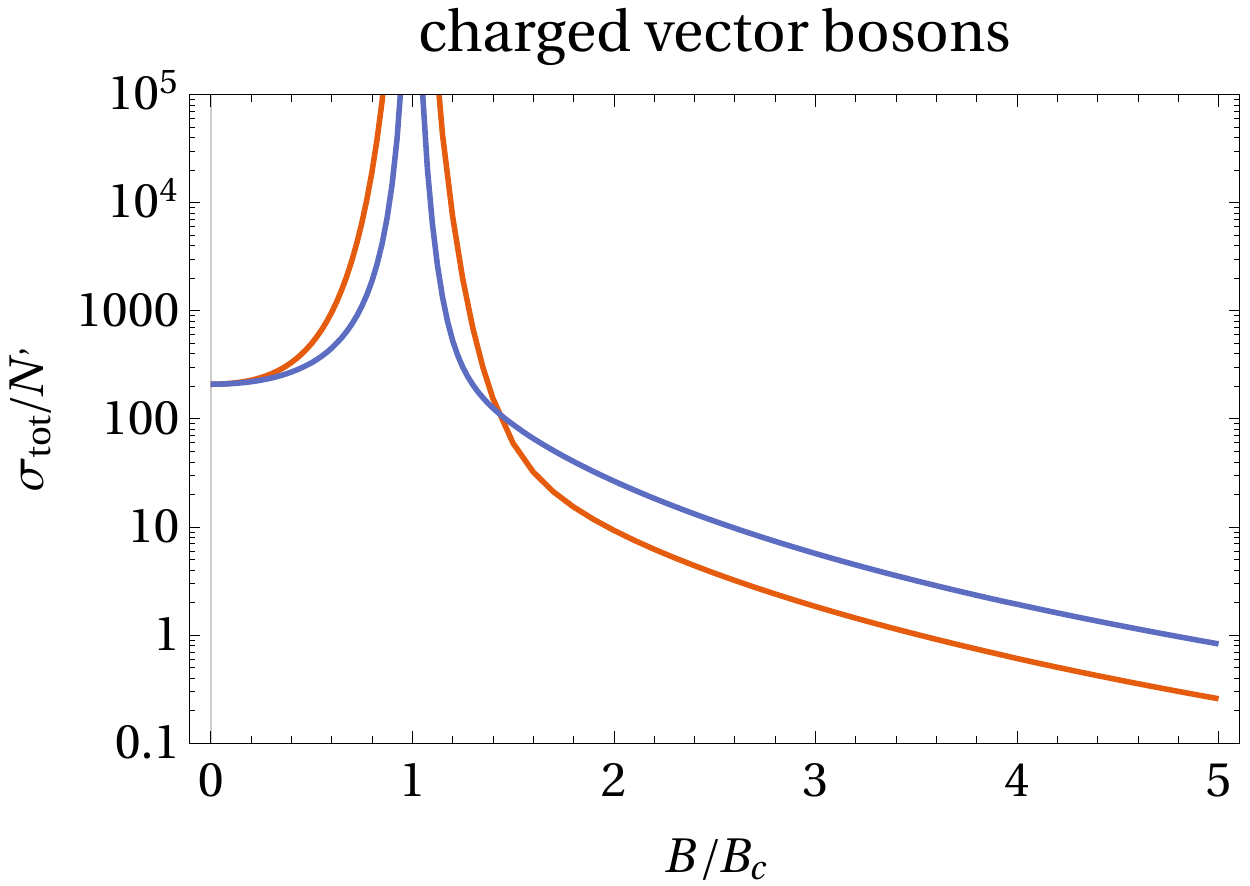}
\caption{\label{fig:stotvectorlog}
Total unpolarized light-by-light scattering cross section for virtual charged vector bosons with $g=2$
as a function of $\xi=B/B_c$
with the magnetic field perpendicular (red lines) and longitudinal (blue lines) to the collision axis,
normalized as in Fig.~\ref{fig:stotvector}, but now plotted up to $\xi=5$ in logarithmic scale.
At the critical field strength, $B=B_c$, the cross section diverges and perturbation theory breaks down. 
The latter is also the case for $B>B_c$, as the Lagrangian density
has an imaginary part, signaling instability against vector meson condensation.
}
\end{figure}

In the light-by-light scattering cross section as derived from the Euler-Heisenberg Lagrangian, 
the vanishing of the effective mass in the lowest Landau level
leads to a divergence, shown in Fig.~\ref{fig:stotvectorlog}, indicating a breakdown of perturbation theory.
Indeed, the range of validity of the calculation changes from $\omega\ll m$ to $\omega\ll m_{\rm eff}$,
i.e.\ $\omega^2/m^2 \ll 1-\xi$, for charged vector bosons with $g=2$.

The divergence of the light-by-light scattering amplitude caused by charged vector bosons can be traced to
the spin contribution in (\ref{Lchargedvector}). Expanding the integrand on the right-hand side of (\ref{Lchargedvector})
in powers of $b$, the integral can be evaluated with result
\bea
\Lc_{\text{vector}} - 3 \Lc_{\text{scalar}} &=&
\frac1{8\pi^2}
\left[ 2 (ea)^2- (ea)^2 \ln\left(1-\frac{(ea)^2}{m^4}\right)
+ea m^2 \ln\frac{m^2-ea}{m^2+ea}
\right]\nonumber\\
&&+\frac{(ea)^2}{24\pi(m^4-(ea)^2)}(eb)^2
+\frac{7(ea)^2(3m^4+(ea)^2)}{720\pi^2(m^4-(ea)^2)^3}(eb)^4+O(b^6),\quad
\eea
where $m^2$ is to be understood as having an infinitesimal negative imaginary part, $m^2\to m^2-i\epsilon$,
when $ea\ge m^2$. Evidently, there is a singularity at $ea=m^2$ which leads to a multiple pole in the scattering
amplitude at $B=B_c$. For $ea>m^2$, a finite result is obtained, but the Lagrangian then has an imaginary
part at $b=0$, i.e., for a purely magnetic background field, which corresponds to the possibility \cite{Chernodub:2010qx,Hidaka:2012mz,Chernodub:2013uja} of
the decay of the vacuum into a superconducting state of condensed charged vector bosons.

\section{Polar diagrams for unpolarized cross sections}
\label{sec:polardiagrams}

\begin{figure}
\includegraphics[width=0.5\textwidth]{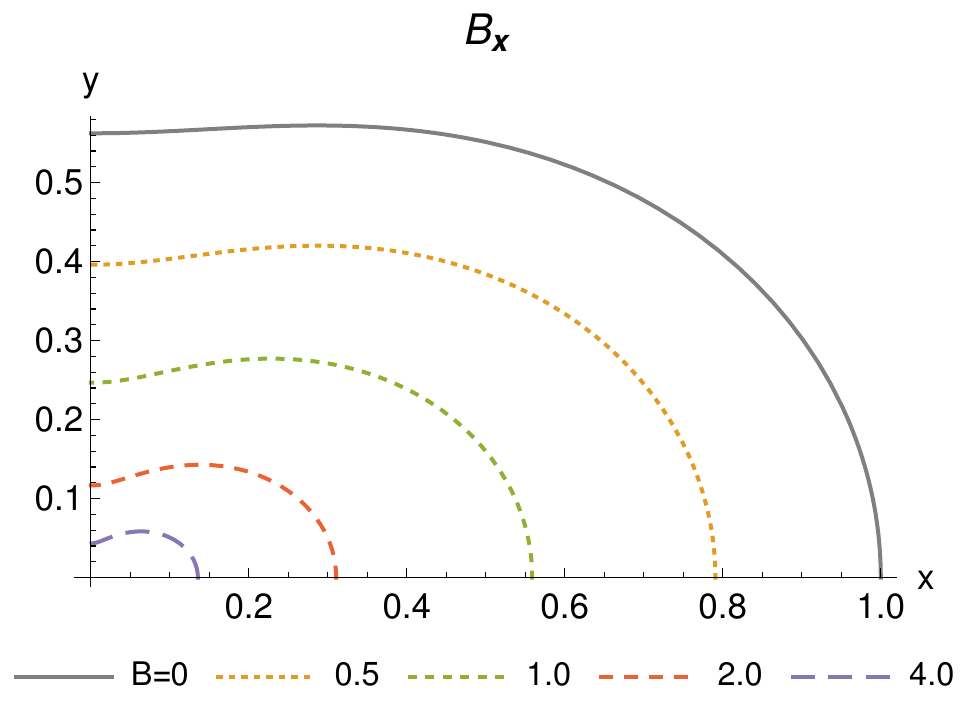}
\includegraphics[width=0.5\textwidth]{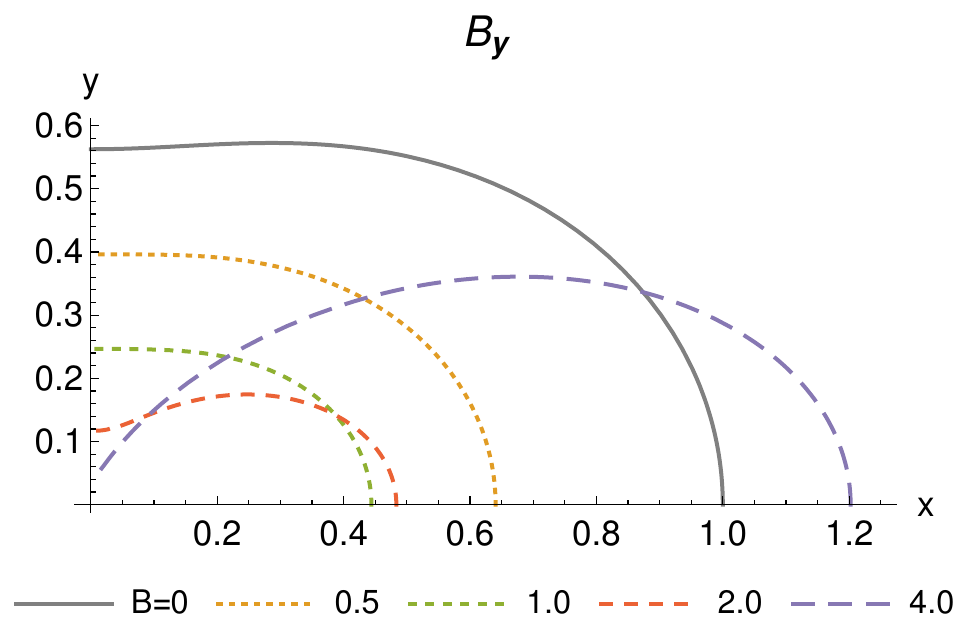}
\includegraphics[width=0.5\textwidth]{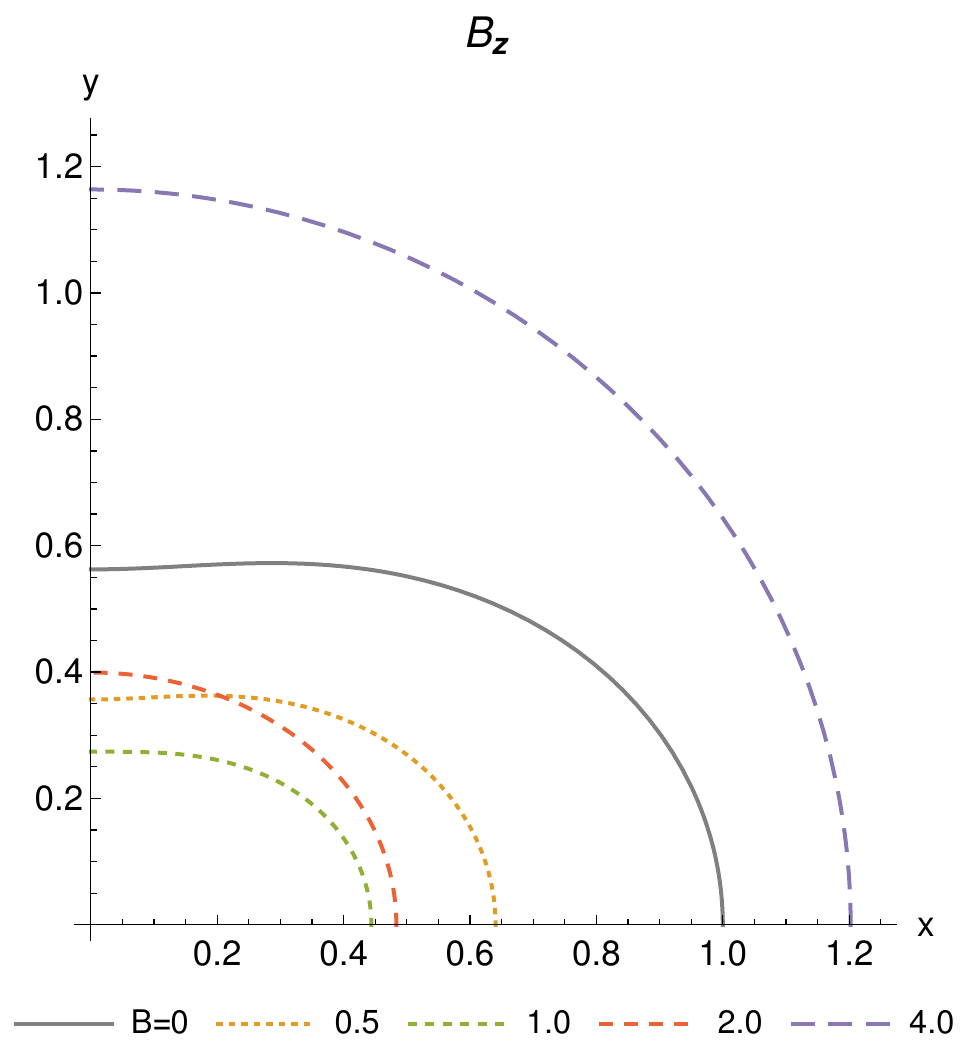}
\caption{\label{fig:polar}Polar diagrams of the unpolarized cross section $d\sigma/d\Omega$ in the scattering plane
for spinor QED with various background magnetic field strength (in units of $B_c$)
and three orientations of the magnetic field (coordinates as in Fig.~\ref{fig:kinematics}, cross section normalized to forward scattering
at $B=0$). Top: $\VB$ in $x$-direction, i.e., parallel to the collision axis;
center: $\VB$ in $y$-direction, in the scattering plane and orthogonal to the collision axis; bottom: $\VB$ in $z$-direction, orthogonal to the scattering plane. (Only one quadrant of the polar plot is shown.)}
\end{figure}

In Fig.~\ref{fig:polar} we display the unpolarized differential cross section of spinor QED\footnote{The corresponding diagrams for scalar QED show less structure; the cross section
reduces rapidly as a function of $B/B_c$ in all directions. For charged massive vector bosons, the cross section
increases in all directions as long as $B<B_c$.} as a function of the strength of the background
magnetic field for three orientations of the magnetic field with respect to the scattering plane (chosen as the $xy$-plane,
see Fig.~\ref{fig:kinematics}). For small to medium field strength, the cross section decreases with $B/B_c$ in all directions, but
at high field strength it rises again in directions orthogonal to $\VB$.

\newpage
\bibliographystyle{JHEP}
\bibliography{lblmagnetic}
\end{document}